# Usability Testing of an Explainable AI-enhanced Tool for Clinical Decision Support: Insights from the Reflexive Thematic Analysis

Mohammad Golam Kibria, PhD, MPH, MBA
*Maternal Fetal Medicine*
*University of North Carolina at Chapel Hill*
*Pediatrics Department, Wake Forest University School of Medicine*
North Carolina, USA
kibria@email.unc.edu

Alison Stuebe, MD, MSc
*Maternal Fetal Medicine*
*Maternal and Child Health, Gillings School of Global Public Health*
*University of North Carolina at Chapel Hill*
North Carolina, USA
alison_stuebe@med.unc.edu

Lauren Kucirka, MD, PhD
*Maternal Fetal Medicine*
*School of Data Science and Society*
*University of North Carolina at Chapel Hill*
North Carolina, USA
lauren_kucirka@med.unc.edu

Javed Mostafa, PhD
*Faculty of Information*
*University of Toronto*
Toronto, Canada
dr.javedm@utoronto.ca

***Abstract*—Artificial intelligence-augmented technology represents a considerable opportunity for improving healthcare delivery. Significant progress has been made to demonstrate the value of complex models to enhance clinicians' efficiency in decision-making. However, the clinical adoption of such models is limited due to multifaceted implementation issues. One documented area of concern is the unclear AI explainability that negatively influences clinicians' acceptance of a complex model. With a usability study engaging 20 U.S.-based clinicians and following the qualitative reflexive thematic analysis, this study develops and presents a concrete framework and an operational definition of explainability in AI. The framework can inform the required customizations and feature developments in AI tools to support clinicians' preferences and enhance their acceptance.**

## I. INTRODUCTION

Despite the significant potential of Machine Learning (ML)-driven Artificial Intelligence (AI) tools for clinicians to enhance patient care and healthcare delivery, their integration into clinical care faces challenges. A plethora of research demonstrates the promise of ML models for healthcare delivery, particularly in areas like postpartum depression [1], [2]. However, a significant gap still exists in translating these models into practical risk prediction tools with only sporadic implementation observed. For example, in postpartum depression care, delayed adoption of such tools within electronic health records system have been noted [3]. Issues with model opaqueness, complexity and usability of these tools likely create hurdles to acceptance [4]. Understanding clinicians' perceptions, expectations, and concerns about model accuracy, bias, and explainability is critical to identify barriers to use [5], [6].

In simple terms, explainability refers to the ability of an AI system to present the rationale of its decision-making processes, which typically is missing in many AI-anchored decision support systems and create loss of confidence among users and becomes a barrier to broad adoption. The need for algorithmic accountability has led the European Union to enact the General Data Protection Regulation (GDPR) [7]. This law requires the "right to explanation", which mandates AI systems provide the explanations behind their decision-making process if patients ask for it [7]. However, it is critical that clinicians also understand the output and that it is delivered with information they prefer to see. AI researchers have made significant efforts in the past few years to develop several explainable AI (XAI) methods to tackle this implementation challenge [8]. These methods have the potential of engendering confidence, particularly in complex models, often referred to as black-box models, and increase their adoption and use by clinicians.

*Research problems, aims and questions*

A limited number of prior studies has demonstrated the power of explainability in AI-anchored systems in certain non-healthcare domains, however, there is a significant lag in incorporating and adopting XAI in healthcare tools and platforms [9]. Little is known about clinicians' explainability needs and their perspectives towards XAI-based healthcare systems. Furthermore, it is unclear what makes a system explainable and what dimensions may improve user acceptance, and there is a lack of standardized and well accepted definitions of explainability in existing literature. For example, researchers identified 36 notions in a systematic review that contributes to the construct of explainability, which might be overwhelming to the busy clinicians [10]. Additionally, there are no formal constructs to determine whether improved explainability actually enhances understanding, as oversimplified models could be perceived as highly explainable but may be less accurate and can potentially misguide clinicians. As such, this research aimed to identify the underlying constructs and to develop a formal definition of explainability and framework for evaluation through a study directly engaging with clinicians.

This study extends the previous research on the quantitative aspect of the usability study for assessing AI explainability (*citation omitted due to double-blind review*). This paper incorporates the qualitative data of the study and provides a more in-depth analysis of the underlying dimensions of explainability framework. The qualitative approach was also employed as it is ideal for expanding and deepening researchers' understanding and exploring new themes from such rich data collected from clinicians. A diverse group of clinicians varied clinical specialties were recruited to interact with an AI prototype and collected their verbal and written feedback related to their perceptions and expectations of the tool. This study bridges the research gap between Human-Computer Interaction (HCI) and explainability involving clinicians for interacting with an explainable AI prototype, aligning the recent findings that <1% of XAI papers validate explainability with humans [11]. In this study, the feature-based SHapley Additive exPlanation-SHAP interpretability method was employed, which is the most widely accepted explanation approach in healthcare [12], [13]. The research poses several questions that include:

i) Can a qualitative usability study engaging clinicians uncover the dimensions of AI explainability?
ii) Can clear definitions of AI explainability be established based on the identified dimensions?

The subsequent sections present the methods and materials, including study settings, participant details, a brief overview of the qualitative reflexive thematic analysis and the set-up for usability testing of an XAI tool that predicts postpartum depression as a use case study. The result section presents clinicians' expectations, perceptions and level of satisfactions regarding the tool. Following that, a compilation of clinicians' input and feedback is also presented to identify and define the key themes that influence explainability of AI tools. Next, a brief discussion addresses the study findings and their implications, as well as linking them with existing literature. Finally, the conclusion section summarizes the lesson learned, strengths, limitations of the study and the broader implications of the results for future research directions.

## II. METHODOLOGY AND EVALUATION PLAN

### *A. Study setting, participants and recruitment*

This study recruited 20 clinicians (maternal and fetal medicine, general obstetrics and gynecology, resident physicians, psychiatrists, nurse practitioners and midwives (certified professional midwife/licensed midwife) who reside in the US, aged 18 or older, were fluent in English, and able to participate in in-person or online sessions. The sample sizes employed in this study were considered adequate from both qualitative approach and usability study perspectives. With 20 study participants, the key themes were identified within the suggested ranges for theoretical saturation [14]. Empirical evidence also demonstrated that such a sample size would sufficiently reveal the majority of usability issues [15]. Multiple recruitment strategies were employed including flyers, purposive, and snowball sampling methods.

### *B. Study procedures*

The lead researcher established a communication channel with the participants who showed interest via email for arranging the session. A 45-minute remote usability study was conducted with clinicians via web conference tool (Zoom) between July and October 2024. The session was audio and video recorded to better understand their expectations, perceptions, and satisfaction with AI tools, allowing for better understanding without repeat interviews. No prior training or user manual were provided to participants to ensure unbiased user interactions. Participants did not receive compensations for participating in the study.

### *C. The postpartum depression AI tool*

To effectively execute this study, researchers developed an AI-driven Postpartum Depression (PPD) tool, to assess postpartum depression risk score. The tool provided personalized risk assessments of pregnant individuals and presented interactive explanations of how the AI system arrived at its decisions. The PPD tool provided details on project objectives, data source, and information related to sample size, pre-processing, data resampling approach, class balancing, cross validation, hyper parameter tuning, and performance metrics including ROC_AUC, sensitivity, specificity etc. of the selected model (*citation omitted for double-blind review*). The online tool featured a calculator interface to provide patient-specific inputs and generate individualized-level risk and risk score for developing PPD. The tool also visually presented the rationale behind the AI tool's conclusions, utilizing the explainable local model - SHAP, that illustrated the features contributing to or protecting from risks [16]. The PPD tool was designed, iterated and improved in consultation and collaboration with health informaticians and clinicians and prioritized individual's clinical responsibilities, cognitive load and time needed to interpret the results in the design principles.

The results were presented both in percentage risk (probability score toward risk) and with a color-coded gauge chart (green for low risk (<=50%, i.e. there is less than a 50% risk that the patient will develop depression in the postpartum period), orange for medium risk (>50 and <=70), and red for high risk (>70%).

### *D. Illustration of a use case*

With the user input information shown in Fig. 1, a user received the risk score and subsequently patient-specific feedback (Fig. 2). This hypothetical patient information shown below was provided to demonstrate the features and functionalities of the tool.

***maternal age in category****: <=17;* ***maternal race****: Black;* ***paternal-race White****: No;* ***pregnancy intention****: Later;* ***Depression during last pregnancy****: Yes;* ***Tooth clean:*** *No;* ***Insurance paid by Medicaid****: Yes.*

*In response, the PPD tool's prediction output was:*
*Risk score = 74.6%, which is classified as "**high risk**" considering the pre-determined threshold, with the blue arc progressed towards the "**red**" zone.*

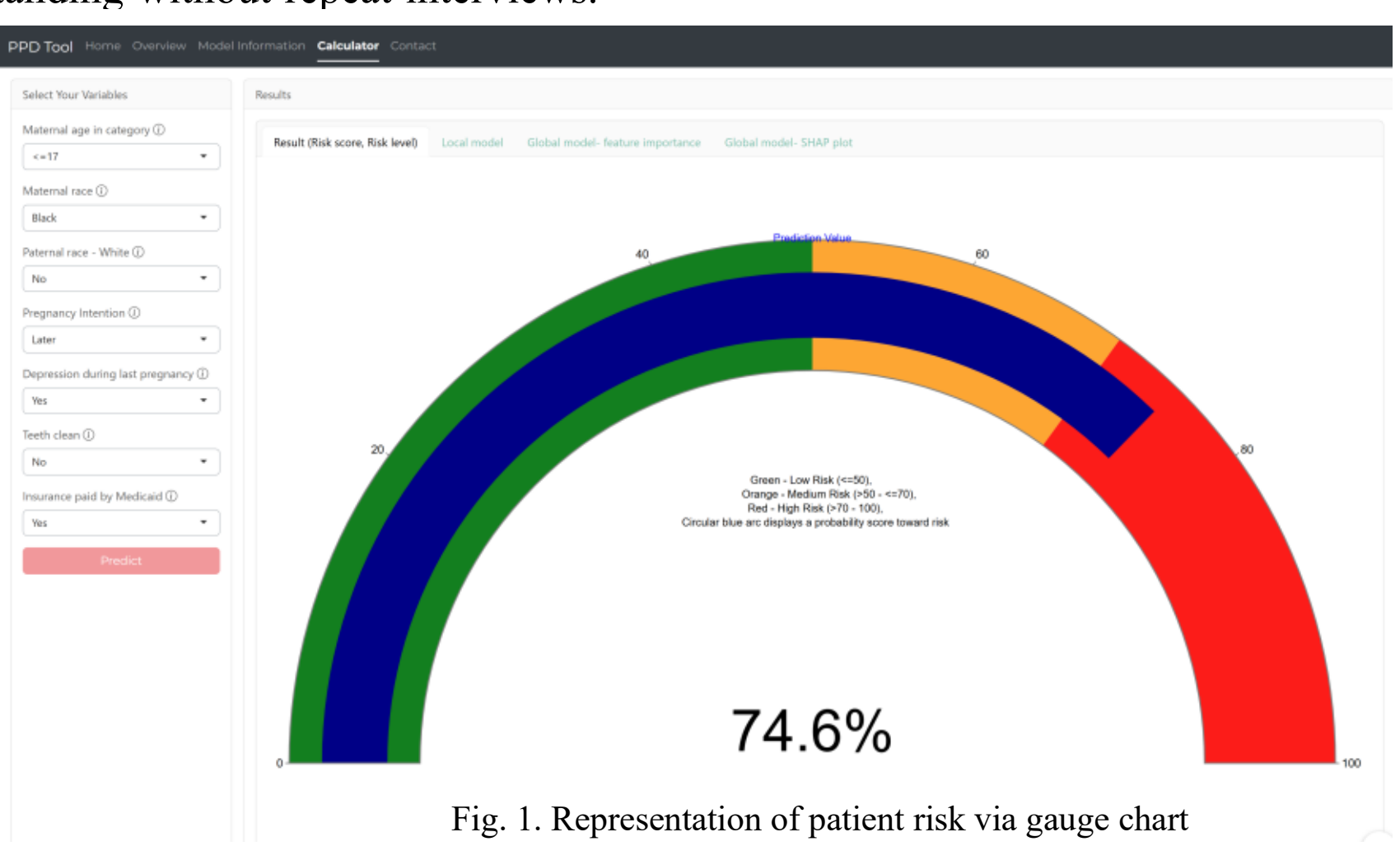

Fig. 1. Representation of patient risk via gauge chart

In Fig. 2, the user gained insights into the input features that affected an individual's postpartum depression risk. In this example, the history of depression, mother's age, and use of Medicaid were the top three contributors toward being a high-risk patient as illustrated in the SHAP plot (a short note on the mathematical properties of SHAP is provided at the end of this article).

Data was used from Pregnancy Risk Assessment Monitoring System (PRAMS), administered by U.S. Centers for Disease Control and Prevention (CDC) [17], and implemented a "black box" machine learning algorithm (i.e. XGBoost). Access to this online tool (https://ppd.lairhub.com/) required login credentials since this current version was only for academic purposes and was accessible upon request.

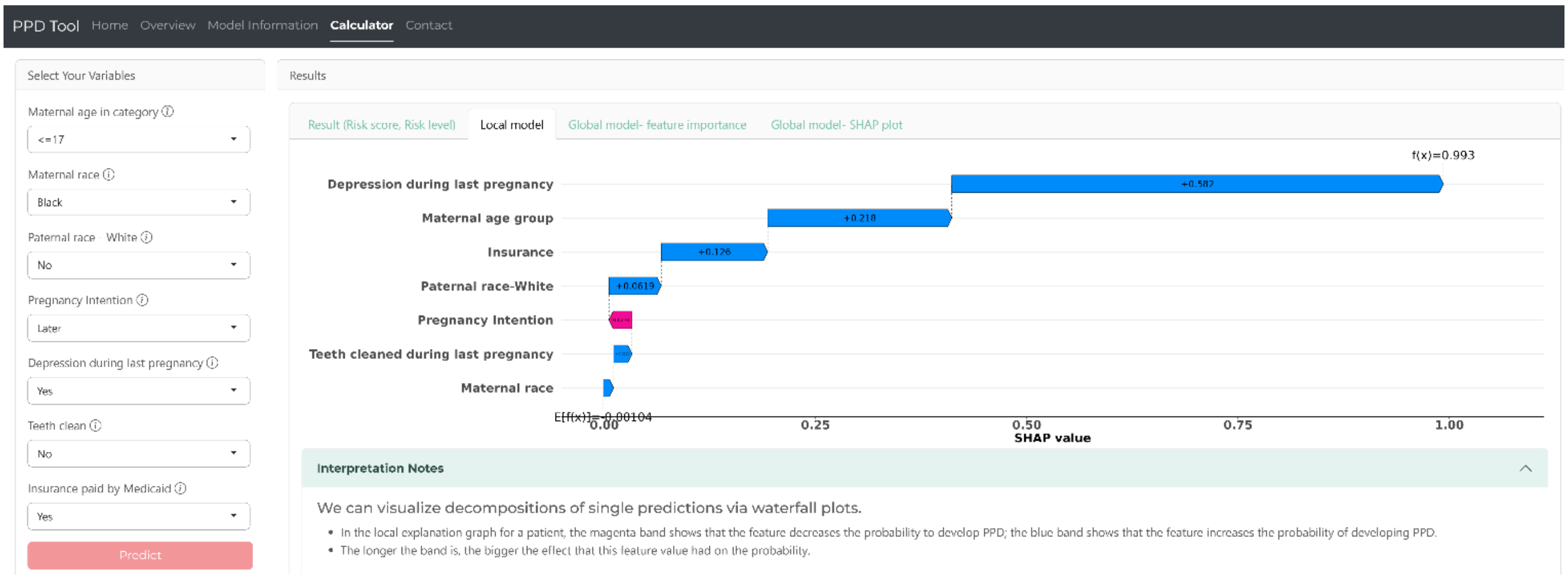

Fig. 2. Representation of features that contribute or mitigates to risk (local model)

*E. Qualitative data synthesis*

Following the demonstration of the PPD tool above, the subsequent step involved explaining the qualitative methodology employed in this paper. The reflexive thematic analysis (RTA) framework was utilized to analyze the clinicians' comments from the usability study. RTA is an accessible and theoretically flexible interpretative approach since it facilitates the identification of themes, conducted at the intersection of: (1) the dataset; (2) the theoretical assumptions of the analysis, and (3) the analytical skills/resources of the researcher [18], [19]. The study adopted a fundamental principle that reflects clinicians' belief and encounter with AI system, professional experiences and clinical reasoning. Based on the nature of this usability study, it was posited that the RTA is the most suitable framework within this context. The qualitative results in the user interaction sessions were undertaken to identify AI explainability-related constructs to the adoption of AI-driven tools in healthcare, using a six phases of reflexive thematic analysis method [18]. The phases include: (1) familiarization with the data, (2) initial code generation, (3) initial theme generation, (4) reviewing and developing themes, (5) Refining, defining and naming themes, and (6) producing the report.

In this qualitative thematic analysis, a collaborative and reflexive methodology was employed by involving multiple coders to critically evaluate the concepts and interpretations derived from the data and build consensus. The primary coder (health informatician) consistently updated the code through multiple iterations following the RTA. The secondary coder with a background of MD and PhD independently coded the full dataset to check the validity of the coding. An inter-rater reliability test was conducted. This methodological approach was facilitated to achieve nuanced interpretations, concurrently fostering consensus building on the identified themes [19], [20]. All aspects of the emergent themes and interconnections that cohere around the central concept of explainability were considered. See the process of theme developments as an illustrative example in table 1.

TABLE 1. THEMATIC ANALYSIS PROCESS

| Participant Number | Age | Comments | Data Extract | Open coding | Axial coding | Initial Theme | Final Theme |
| --- | --- | --- | --- | --- | --- | --- | --- |
| **Participant 18** | 31 | *My one thing is that makes me a little wary of using machine learning algorithms is that there's evidence, and some other uses of this, that because it's learning from human data in the US, it's like learned to make biased decisions.* | Concerned with AI-bias in decision making | Concerned with algorithm bias in general | Perceived bias and risk of biased decisions | Trustworthiness in ML model's decision due to inherent bias | Trust |

*1) Usability Testing*

To study the effect on clinicians' implementation of the AI tool, a usability testing was employed as mentioned earlier, which had two phases – 1) pre-study questionnaire and 2) users' interaction session with the tool following a set of tasks. Both positive impressions and unmet needs were gathered regarding the tool. The pre-study questionnaire contained a list of multiple-choice (response required) and short answers (response required) to understand participants' background, experience, relevant expertise, and expectations from an AI-driven risk assessment tool. Participants were given a set of assignments aimed at performing specific scenario-based tasks. For example, they were assigned to create a hypothetical patient and enter the data into the tool and then generate individualized risk score. Their interactions with the tool were observed to identify usability issues and were asked to interpret the prediction results using patient-specific information outputs and visualizations. During the summative usability test, feedback was collected from the clinicians through the "Think Aloud" approach while they interacted with the tool [21]. In this method of usability testing, the users verbalize their thoughts while interacting with the tool and provide comments based on their views in an unrestrained manner. This method is considered as the "gold standard" for usability evaluation [22].

*II) Data collection and analysis*

The findings derived from individual pre-study interviews and user interaction sessions were summarized and presented in this paper. The online Qualtrics tool was used for pre-study data collection and the recordings were stored on a protected drive [23]. The recording data was transcribed using Zoom transcription software and then checked and edited as appropriate to correct errors and inaccuracies. The data was then converted into an anonymous format for analysis, guided by team members with expertise in similar qualitative methods in usability testing in healthcare. The extracted data was tabulated in Microsoft Excel (Microsoft, Redmond, WA, USA), according to the study aim. All analyses were performed using the R version 4.4.2, and significance was defined as <0.05 [24].

*F. IRB Approval*

In all cases, the lead investigator recorded participants' responses. The study was approved by the Institutional Review Board (IRB). All participants gave written consent prior to the study.

## III. RESULTS

*A. Participants demographics*

This usability study engaged 20 clinicians, 14 (70%) were female and six (30%) were male. Among them, 11 (55%) had MD degrees, six (30%) had nursing degrees, and three (15%) were mid-wives. The average age of participant was 37.35 years (standard deviation = 7.06 years) with an average of 10.30 years (standard deviation = 6.17 years) of professional experience. Only three clinicians had prior experience using AI/ML-driven risk prediction tools; one found it very effective, while the other assessed it as having minimal to acceptable level impact.

*B. Session duration*

All the participants, who are typical clinical end-user engaged with all six critical components of the system, namely the overview page, model information page and four pages associated with calculator itself, represented as tabs on the main window in the tool and completed 100% of the assigned tasks. As the sample size is below 25, the "geometric mean" for interaction time calculation, minimizing potential error and bias was employed [25]. The average (geometric) interaction time for clinicians was 1,193 seconds (95% CI: 958 -1,484 seconds). Among 20 clinicians, eight entered data for more than three patients to explore hypothetical scenarios and generate risk predictions. This activity helped them understand the model behavior in risk prediction and facilitated interpretation of the local SHAP model to clarify the contributing or mitigating features. This exercise helped assess their trust and confidence in the model and the outputs. Clinicians required a median of 50 seconds to input the first patient data, followed by 25 seconds for the second and 26 seconds for the third, culminating in prediction results.

*C. Thematic analysis*

The RTA approach of the usability study in both pre-study questionnaire and the users' interaction session resulted in four key themes critical to explainability, which include the dimensions of 1) understandability, 2) trust, 3) usability, and 4) usefulness. The thematic dimensions were identified based on the information and comments provided by the users in the pre-study questionnaire and were consistently reflected in the user interaction sessions ("Think Aloud" protocol).

*1) Pre-study questionnaire (interview) session*

In the pre-study interview phase, clinicians were asked one of the open-ended questions: "*Which aspects of the risk assessment tool do you think should be prioritized during development*?". This question aimed to gather their insights and expectations regarding an AI tool before they use it. Clinicians' comments were systematically open coded following the RTA process by two researchers, summarized and grouped into similar themes and tallied to identify recurring expectations. The frequency of each coded items directly informs the thematic distributions as displayed in Fig 3, which is helpful to understand clinicians' perspectives.

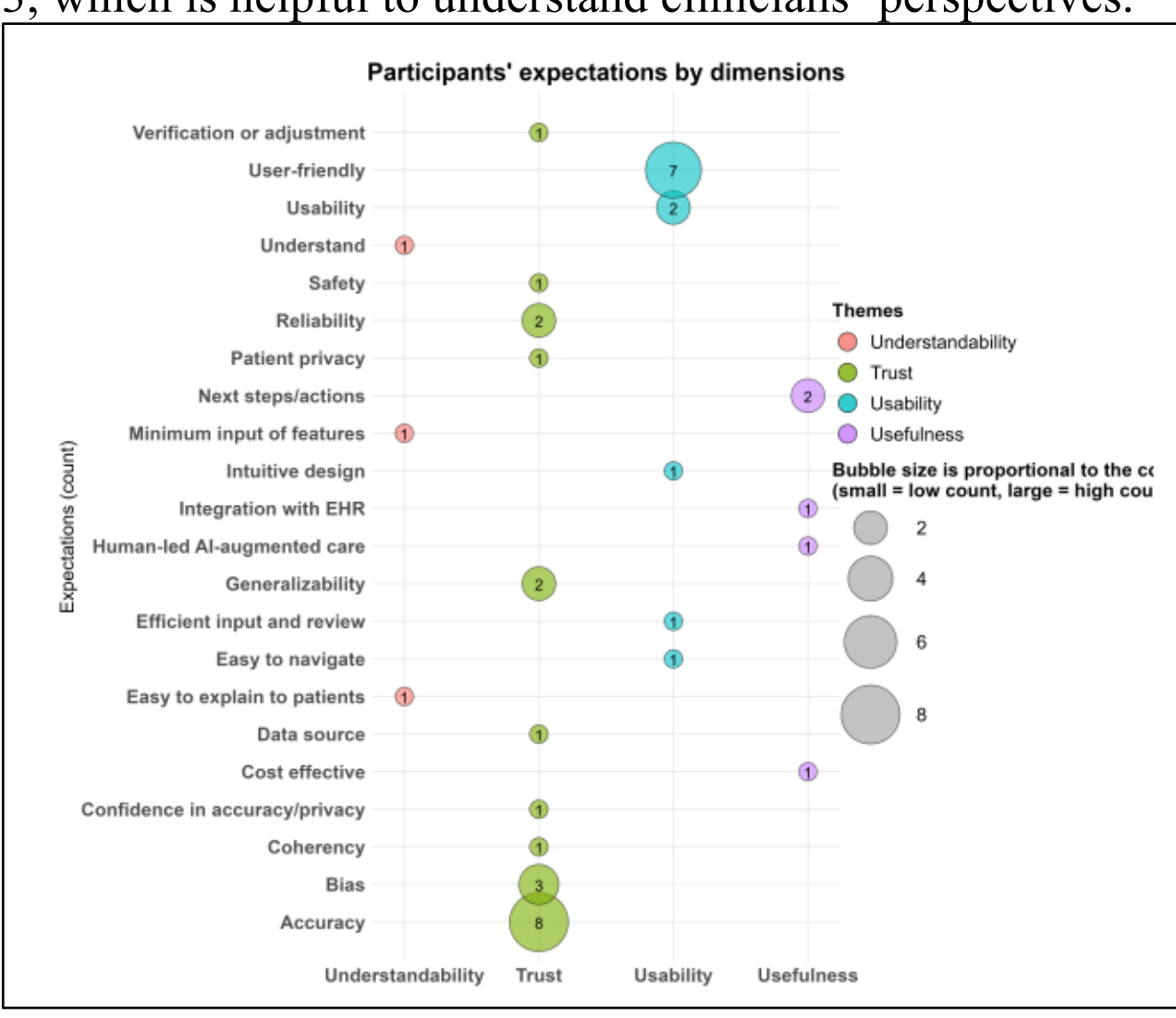

Fig. 3. Pre-study questionnaire: user expectations of an AI tool

*2) Usability task – user interaction (testing) session*

The feedback collected from the users was also reviewed and analyzed. The agreement between the two coders was analyzed using both percentage agreement and Cohen's Kappa in the inter-rater reliability test. A total of 68 themes were identified, and they were grouped into four unique recurring themes. The percentage agreement was 78% (95% CI: 66.24 – 87.09), indicating that the coders assigned the same themes to 78% of the cases. Cohen's Kappa was also implemented to determine the agreement between the coders: $\kappa = 0.69$ (95% CI: 0.56–0.83), $p < 0.0001$. A Kappa score greater than 0.60 reflects a substantial agreement between the two coders' judgements. The main themes that emerged from qualitative data that fall broadly within the AI explainability context were understandability, trust, usability, and usefulness. The summary of key themes identified in the user interaction (testing) sessions is accompanied by overarching concepts, and representative clinician quotes are presented below for clarity and evidence.

***Representative quotes from the clinicians:***

**Theme: Understandability**

Clinicians reported a varying level of ease regarding their understanding of the AI tool. They felt experienced users would likely perceive such tools as easy and intuitive, whereas novices might struggle.

- *That's another thing I thought about, too, like this is really easy, like I can do all this. This is nothing to me, because I have a lot of experience. But the answers to these the average person might not agree with, or they don't have the background. Knowledge.* ***Participant 1, age-29***

However, they suggested that their comprehension skills could be improved through continuous use and learning. They felt more interactions with the tool would improve their ability to understand complex methods and graphs, like global SHAP model plot.

- *That's so interesting. And I could see being a care provider ...you are learning from the tool. So, like the more you use the tool, the more you begin to know what factors are going to be protective or a risk level. Like the more you use it.*
  ***(Participant 6, age-41)***

Participants appreciated the representation of color-coded risk metric; however, one participant cautioned against the potential of misinterpretation, particularly among novice professionals, due to an oversimplified (cut-off value) representation of the risk. The participant advocated for the reevaluation of risk threshold value based on scientific evidence to bolster transparency.

- *I like that. You have the information in terms of how the model is developed…What a clinician would most likely be using would be this piece of things. I would really want to make sure there was a like a clear, scientific way in which these green, orange and red zones were developed, and that that was based on something that was truly actionable as opposed to just arbitrary cut off (gauge chart to present the risk metrics). Be very concerned, because, you know, somebody who's an early level clinician will decide that yellow. This yellow is totally fine, whereas this red is not. And if that's not actually true, I would be very worried about people utilizing, especially these types of cut points in that way.* ***(Participant 16, age-43)***

**Theme: Trust**

Clinicians expressed generally a positive impression of the predictive accuracy of the tool, thereby enhancing their confidence in the output generated by the model.

- *My hesitancy would always be things like accuracy and privacy. Bias. So, I would love to have this tool. Oh, I think it's fantastic.* ***(Participant 3, age-53)***
- *I guess so, because CDC data set, you know, it's still even that… it's not without bias. I was confident in the reliability in the various scenarios.* ***(Participant 16, age-43)***

However, they highlighted the necessity of personal validation by comparing the tool's accuracy with their own assessments. Participants shared their perceptions about algorithm bias generally seen in AI tools, which potentially might lead to biased decisions.

- *I'm like I'm super curious to see how it comes out, and compares to some of the other ones, or just like my own interpretation, like my own evaluations of people just from. You know, experience and knowing and relationship to them. That'll be kind of fun to like. See how that compares.* ***(Participant 7, age-43)***
- *My one thing is that makes me a little wary of using machine learning algorithms is that there's evidence, and some other uses of this, that because it's learning from human data in the US, it's like learned to make biased decisions.* ***(Participant 18, age-31)***

Largely, the participants also stressed that ensuring patient privacy and security standards would foster their trust.

**Theme: Usability**

Clinicians underscored the usability strengths of the AI tool, emphasizing its intuitive design, easy navigation, lucid feature representation, and the simplicity of interpreting its visualizations. They also appreciated the interactive nature of the tool, particularly the dynamic visualization of feature contributions, thereby augmenting user comprehension. The tool's real-time responsiveness and ability to present feature weightage values were recognized as critical features that improved usability experience and decision-making capacity.

- *One of the things that I love about this feature is that it's easier to navigate. The features are easier to understand. And they are clear.* ***(Participant 2, age-28)***
- *I think it's cool, because I like the dynamic way you can change the different factors. It's like a visual tool that you can apply. The fact that it pops up and it changes*

- *immediately. It computes very well. Like it's easy for me to understand what's happened. Yeah, because it shows you which way the factor is like pushing the arrow (local SHAP model). This one will help me.* ***(Participant 20, age-31)***
- *I like the way these data are set up. I like the fact that you can see exactly which factor and how much it's contributing to the eligible prediction score. I do like that piece. yeah, like, I said, I, I like that. It is color-coded in terms of things that increase the risk versus decrease, the risk. I like that. It goes in order of what's most important. I like that.* ***(Participant 16, age-43)***

Participants also valued the tool's ability to present patients' personalized risks through local models and to explore global models, aiding their decision making. The color-coded risk representation and well-labelled features with tooltips facilitated user understanding.

- *I don't know if it's possible to look at the data (raw data). Well, no, because you have it in the local model, right? The local model has got all the breakdowns of the groups. Yeah. So that's good, too. Good that it's interactive cause. I think, too, that being able to witness, like almost live, if you will, to see how the different factors affect it are great for someone to understand the impact. Everything was well labeled, too. So, it was easy, like, I feel like, if you sent this to somebody who you didn't explain it to. They could understand where to navigate. The variables are succinct and getting to the point. Cut and dry. Interpretation notes are helpful. Interactive (live feed) impacts on knowledge.* ***(Participant 14, age-29)***

However, they suggested improving the navigation feature by incorporating direct links between related screens. While participants found interpretation notes for the graphics to be useful, they underscored the incorporation of additional contextual explanations, which could further assist users with varying levels of expertise.

- *So, I think, because of the flow … I see this screen, then I should go to another screen to see more information. I think maybe you can have… like a button here to navigate. Or to the end add to the next screen, to next screen, to see more information on that.* ***(Participant 15, age-35)***
- *I feel … like as a participant who understands research, like, I can you know interpret these graphs better. As someone who may not have as high of an education level. You know they might. And like it still might be interesting to them to look. But I mean, I think the overall result risk score is one that's most helpful. Right? I would just I would have them all there. Make the risk score prominent (tab/highlight). And obvious. You know what - Raw data view. We'll understand, you know. And what they can do.* ***(Participant 13, age-38)***

**Theme: Usefulness**

Clinicians highlighted the potential utility of AI driven tools in both clinical decision-making and the provision of patient education. While this AI tool's ability to predict patients' risk was appreciated, participants strongly emphasized the necessity for generating actionable recommendations. Participants suggested expanding the tool's functionality into a patient-facing application with the aim of improving patient engagement and fostering self-management practices. Participant 4 and Participant 13 explained their expectations as follows:

- *Once I know, high risk is kind of understandable. But what do I do with that information? What is the next step after we get that? What is the recommendation for the lowest patients?* ***(Participant 4, age-40)***
- *What can you do to with that knowledge? Oh, my score is in the seventies or eighties. What can I do now? It could be a (patient-facing) app as well.* ***(Participant 13, age-38)***

They underscored the broader applicability of such tools to other health conditions and utility in risk assessment, intervention planning and allocation of social services based on the risk scores. See below for a sample of participant feedback.

- *And you could use these quick tools, these quick things that this tool you could even use, this maybe kind of model. It seems like it could be helpful towards the future for other kinds of concepts that are really common, like one that we're doing now is postpartum depression.* ***(Participant 1, age-29)***
- *That I feel is so important - that automated piece is not only the referrals, but also like direct patient education, because many people, especially if they're depressed, will never seek help.* ***(Participant 9, age-49)***
- *Think, based on this. Like, social services aid also can be allocated to whom to give the most aid.* ***(Participant 12, age-42)***

Clinicians also highly valued the potential integration of such tools into the EMR system for facilitating streamlined clinical workflow.

*What I think (is that) most interfaces with AI driven tools that are built into EMR - are built in a way that they already pull the data. Nice, because… It saves your time and everything, you know. When you have a 15 min or 20 min visit. Everything is about … like time.* ***(Participant 17, age-32)***

The distributions of themes identified by participants in pre-study interviews and user testing sessions are presented in Table 2. The summary table showed that only two clinicians (10%) mentioned their expectations related to the concept of understandability in the pre-study interview. However, the frequency of the theme became recurring in the user testing session. 14 participants (70%) discussed this dimension as they actively tried to understand the AI tool's input, process, and outputs. Similarly, while a handful of 10 participants (50%) highlighted the usability-related expectations in the interview phase, 16 (80%) of them provided their perceptions and experiences about the usability aspects after interacting with the tool. The usefulness theme was mentioned by only five participants (25%) in their pre-study interview. However, the number of participants that mentioned usefulness theme increased to eight (40%) in the user testing session. While the number of participants increased from pre-study interviews to testing sessions across all themes except the trust theme. Under this theme, 14 participants (70%) discussed their trust-related concerns in the interview. However, the number was reduced to nine participants (45%) during the user testing session. See the results in Table 2.

**TABLE 2.** HEAT MAP FOR THEME IDENTIFICATIONS (LOW: YELLOW AND HIGH: GREEN)

| Participant # | **Understandability** | | **Trust** | | **Usability** | | **Usefulness** | | Total mentions | # of themes identified |
|---|---|---|---|---|---|---|---|---|---|---|
| | Interview (total mentions = 3) | User testing (total mentions = 17) | Interview (total mentions = 21) | User testing (total mentions = 13) | Interview (total mentions = 12) | User testing (total mentions = 30) | Interview (total mentions = 5) | User testing (total mentions = 8) | | |
| 1 | 0 | 1 | 3 | 1 | 1 | 0 | 1 | 1 | 8 | 4 |
| 2 | 0 | 0 | 0 | 0 | 2 | 4 | 0 | 0 | 6 | 1 |
| 3 | 0 | 1 | 2 | 1 | 0 | 1 | 0 | 0 | 5 | 3 |
| 4 | 2 | 1 | 2 | 0 | 1 | 1 | 0 | 1 | **8** | 4 |
| 5 | 0 | 0 | 2 | 1 | 1 | 1 | 0 | 1 | 6 | 3 |
| 6 | 0 | 3 | 2 | 0 | 0 | 1 | 1 | 1 | **8** | 4 |
| 7 | 1 | 1 | 0 | 1 | 1 | 1 | 0 | 0 | 5 | 3 |
| 8 | 0 | 0 | 1 | 1 | 0 | 1 | 0 | 0 | 3 | 2 |
| 9 | 0 | 1 | 1 | 3 | 2 | 1 | 0 | 1 | **9** | 4 |
| 10 | 0 | 1 | 1 | 0 | 1 | 1 | 0 | 0 | 4 | 3 |
| 11 | 0 | 0 | 0 | 0 | 0 | 2 | 0 | 0 | 2 | 1 |
| 12 | 0 | 1 | 0 | 0 | 0 | 1 | 0 | 1 | 3 | 3 |
| 13 | 0 | 1 | 0 | 0 | 1 | 4 | 0 | 1 | 7 | 3 |
| 14 | 0 | 1 | 2 | 0 | 0 | 2 | 0 | 0 | 5 | 3 |
| 15 | 0 | 0 | 1 | 0 | 0 | 4 | 0 | 0 | 5 | 2 |
| 16 | 0 | 1 | 1 | 3 | 1 | 3 | 1 | 0 | **10** | 4 |
| 17 | 0 | 1 | 1 | 1 | 0 | 0 | 0 | 1 | 4 | 3 |
| 18 | 0 | 0 | 1 | 1 | 0 | 0 | 1 | 0 | 3 | 2 |
| 19 | 0 | 2 | 0 | 0 | 0 | 0 | 0 | 0 | 2 | 1 |
| 20 | 0 | 1 | 1 | 0 | 1 | 2 | 1 | 0 | 6 | 4 |
| # participants mentioned themes | **2** | **14** | **14** | **9** | **10** | **16** | **5** | **8** | | |

Pearson's Chi-squared for Independence test showed that there was an association between study phases (Interview and User testing session) and identification of themes ($\chi^2$ = 14.277, df = 3, p-value = 0.002). Furthermore, when examining combinations of study phases and themes using standardized residuals, it was observed that participants mentioned trust theme in the interview phase significantly more than expected. However, they didn't raise any major concerns related to trust dimension after they used the tool. The reason behind such findings can be attributed to clinicians perceived perspectives to the AI tool. Participants were skeptical about AI tool's accuracy, reliability, and potential for perpetual bias while they were asked to share their viewpoints in the interview session. Their concerns extended to patient privacy and safety, quality of data source, and generalizability of the output. This interesting finding could be related to the fact that the context was different. The pre-study questionnaire was abstract and referred to clinicians' opinions on AI tools in general while the user session was specific and referred to the AI tool they were using. The act of playing with the AI tool in the user interaction session also potentially placed cognitive focus on issues of usability and understandability once trust is established. The observed shift from abstract to interaction-level trust concerns indicate that trust is dynamic and situated. The critical design elements of the prototype, which includes system capability disclosure, provision to provide contextual relevant information, and support user control are consistent with Human-AI interaction design guidelines [26].

In summary, participants consistently prioritized understandability and usability as are essential elements for the interaction quality of AI tools, while trust plays a crucial role in fostering ethical AI (see table 2). This qualitative reflexive thematic analysis also revealed the usefulness dimension, which was not identified in the prior quantitative analysis of this study (*citation omitted for double-blind review*). Participants appreciated the intuitive visualization of patient risk profiles but cautioned that novice healthcare professionals might misinterpret the scores without adequate guidance. They also emphasized the benefits of continuous use and iterative learning to deepen their understanding of the tool's underlying assumptions and algorithm. They may require time and support to fully grasp complex outputs like global SHAP plot. Participants showed confidence in the tool's predictive accuracy, but they emphasized the importance of validating results in clinical settings. They generally showed concerns on the systemic biases that any AI model may perpetuate due to training data and algorithm design. Participants underscored the importance of protecting patient privacy and adhering to security standards. They also highly praised the PPD tool's usability and visualization features and advocated for actionable recommendations to support contextual explanations, data-driven decision making, electronic medical record integration, and development of a patient-facing App to enhance patient engagement and education. Overall, the thematic findings in this study demonstrate that there were four key dimensions, all of which were connected to explainability: **understandability**, **trust**, **usability** and **usefulness**. Based on the above analysis, the following operational definitions for AI explainability were proposed:

**Understandability** of data input, meaning, outputs and decision-making processes.
**Trust** in data security, privacy and **confidence** in the AI methods and their outputs.
**Usability** of the tool, which is exhibited by its seamless interactions and efficient outputs.
**Usefulness** of the tool characterized by its ability to generate personalized actionable recommendations to satisfy the user's information need.

**Explainability** refers to the extent to which clinicians can **understand** the **input data**, interact with the **decision-making** processes, interpret the **outputs and suggestions,** and trust the **predictions** of the AI tool. An explainable AI framework based on the relationships of the dimensions was proposed and illustrated in Fig. 4.

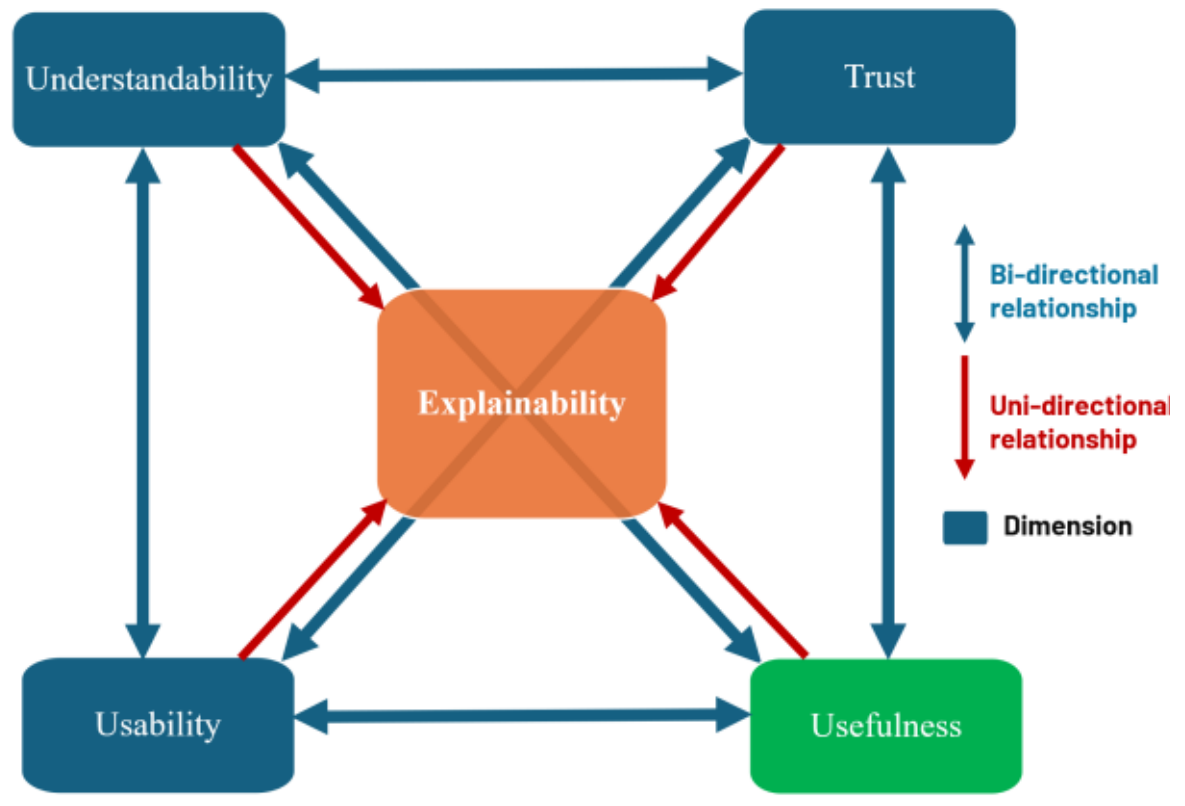

Fig. 4. AI Explainability Framework: connecting four independent dimensions in AI

While explainability and usability are traditionally treated as distinct constructs, this study demonstrates that clinicians primarily experience explainability through the lens of usability, which involves the interaction flow, visual clarity, output evaluation, cognitive alignment with clinical reasoning and opportunities to interrogate the model. These explicit connections clarify why features such as intuitive interface navigation, disclosure of model information, interactive SHAP elements directly influence clinician's understandability and trust. From that perspective, explainability emerges as not only the property of algorithm but as a quality of human-computer interaction.

## IV. DISCUSSION AND LIMITATIONS

The aim of this clinician-participatory usability study was to recognize the fundamental dimensions of AI tools that could serve as a foundation for explainability. Usability studies have proven to be critical in understanding the intricacies of HCI in biomedical research [6], [27]. By involving a pool of clinicians with various backgrounds and employing a qualitative design for data collection, several aspects across four dimensions were detected that may hinder their potential acceptance. The qualitative approach taken by this study provides unique insights and helps collect data from the intended users, providing a depth of knowledge that quantitative methods are often unable to achieve [28]. This study underscores the value of user-focused development over developer-oriented approaches for successful real-world implementation of such tools [29].

The usability study provides researchers with an in depth look at the pilot implementation experience and provides important considerations in advancing the understanding of AI explainability. In this study, clinicians shared in general skepticism on AI tools before they reviewed the PPD tool. However, their feedback related to uncertainty was reduced after using the tool in the testing phase. Although participants did not discuss model understandability, usability, actionability, or usefulness specifically, they provided positive feedback related to all these aspects while interacting with the tool. This highlights the potential for AI driven tools to be more impactful and useful if clinicians are involved in the design phase, provide feedback and have their concerns related to trust addressed. These tools are advancing rapidly and could be integrated into clinical practice. However, regulatory frameworks need to be in place to ensure the effective planning, design and implementation of AI tools. Despite the inclusion of relevant information, such as tooltips and interpretation notes to enhance user's understanding, the study reveals that substantial training is required for clinicians to familiarize themselves with these innovative explainability methods. Our findings indicate that the effectiveness of explanations depends on cognitive load, user expertise and match between explanation artifact and mental models.

While implementing a "details on demand" approach with user accessed detailed information or explanations when required had improved their usability experience. But clinicians still required hand-on training to harness these AI tools effectively. Novice users struggled with the global SHAP model due to its multi-element data visualization complexity which imposed extraneous cognitive load on their working memory. This challenge is consistent with the cognitive load theory, which requires deliberate hands-on training efforts. On the other hand, they found the interactive local SHAP plot more direct, accessible and required less cognitive effort given their shorter time-on-task in real-world situation. The gauge plot and the directionality of SHAP weight were better suited to their mental models and clinical reasoning in the light of feature impact on prompt risk assessment. They felt that the local SHAP analysis empowered them to interrogate the model by changing the input values and observing the resulting changes in predictions, which reinforced their trust in the AI tool. For instance, several clinicians created scenarios by changing the feature values (perturbation) to assess the reliability of the model by creating a patient profile based on their experience with high and low risk patients and expected the model to agree with their clinical knowledge. Expert users were able to triangulate and confirm feature consistency of the local output with the global SHAP model. Such interactions are required to enhance clinicians' understanding of how the model works and to build their confidence. The nature of such interrogation justifies the future relevance and importance of causal prediction approach for building user trust.

This finding highlights the clinicians' involvement in the model building stage, especially feature selection process, which is crucial. Many suggested adding more features. However, clinicians also suggested that the data input should not take much time. The additional feedback from clinicians includes that this type of AI tool study should involve diverse users to get varied perspectives. For example, several participants recommended involving participants without technical backgrounds in research in future studies.

There are several strengths to this study. A sufficient sample size of clinicians was achieved to identify the key themes anchoring explainability and major usability concerns. The study advances existing explainability concepts by demonstrating that clinicians experience explainability not only model transparency, trustworthy, but also actionability of outputs. This dimension was not explicitly operationalized based on user interactions on prior work. The study was conducted online, giving clinicians flexibility in scheduling the session with the lead researcher. This also allowed the study to involve clinicians from several US states. The voluntary interest among the clinicians in interacting with the AI tool, coupled with their transparent and candid feedback, significantly enhanced the insights gained. This clinician-driven evaluation perspective, which is very rare in explainability research provides a solid ground to understand explainability is a clinical AI implementation approach rather than solely a theoretical framework or mathematical model design problem.

The study has several drawbacks – the first being lack of control over the testing environment including information technology equipment and internet bandwidth, which are critical to a satisfactory user experience and uninterrupted human-computer experimental setting. While the themes which were identified through rigorous and in-depth qualitative analysis as critical dimensions for AI explainability, however it requires further validation with more clinicians to fully assess the efficacy of such tools in clinical settings. This study used publicly available dataset for predicting postpartum depression, which might not be the ideal training dataset for building clinician's data-related trust due to concerns for recall bias and self-reporting [30]. Several clinicians highlighted that the risk prediction estimates must be accompanied by a recommendation panel with clinically meaningful actionable insights and guidance for next steps. However, the development of such recommendation panel requires a structured implementation approach involving clinicians, which includes integrating clinical knowledge, treatment protocols, and adherence to regulatory and ethical considerations. EHR-integrated such prompts will enhance the explainability framework by operationalizing the usefulness dimension through bridging the risk estimates to appropriate clinical actions. Although the study identified the dimensions using a single XAI prototype, however, the dimensions are likely transferable to other clinical settings given the widespread use of SHAP-based explanations in modern risk predictions tools.

## V. CONCLUSIONS AND FUTURE DIRECTIONS

Healthcare delivery is dynamic and complex and AI tool is now seen as Software as a Medical Device. Because of this, implementing the AI tools in this domain may be difficult as clinicians' buy-in on the trust dimension upfront is needed. Clinicians are more likely to trust AI tools when they are involved in the model building phase, especially feature selection process so that they can relate to their clinical knowledge and provide feedback. They may be involved in the initial user acceptance testing stage. In the development phase of this type of AI tool study may involve diverse users to get varied perspectives. While model information and resource links help clinicians, they require substantial hands-on training on machine learning models to augment their capacity for model validation.

This study provides a novel four-dimensional framework with clinician derived operational definitions generated through reflexive thematic analysis. This framework reframes explainability construct as interactive, multi-dimensional and action-oriented rather than solely model-centric property. This study expands the existing XAI taxonomies by integrating usefulness as part of explainability, which emphasizes the need for XAI-based real-world clinical applications. Clinicians also suggested an action-oriented interface so that they review the model driven recommendations and can show patients for patient education. This study provides an important contribution to the ethical AI agenda and explainable AI literature by highlighting the necessary customizations needed to support clinicians' preference and enhance their acceptance. Additionally, the usability findings provide insights into how the interactive visualization design can effectively enhance the broader domain of HCI in clinical decision support systems. It is also necessary for the AI tools to demonstrate their utility in the implementation phase. The definition of explainability that emerged from this study captures the underlying dynamics of AI, highlighting its critical role in AI acceptance. The qualitative findings from this reflexive thematic analysis are well-suited within the established implementation science frameworks (e.g., RE-AIM evaluation framework and CFIR determinant framework) by offering clinicians firsthand perspectives on opportunities and barriers towards clinical AI adoption. Additionally, mapping the insights from this study into the HCI models like the Technology Acceptance Model could demonstrate how explainability features can mediate user interactions and influence user adoption in clinical settings. Future work will include adding a recommendation panel to satisfy the usefulness dimension and adopt a suitable implementation science framework to test the efficacy of such framework. Future multi-institutional studies will assess the ecological validity and portability of this framework in an EHR-integrated high-pressure clinical care setting with an expanded list of explanation types with a larger cohort of diverse clinicians. Notably, while the previous quantitative analysis surfaced three core themes related to AI explainability (*citation omitted*), but the qualitative phase revealed an additional dimension–**usefulness**. This unexpected finding underscores the value of reflexive thematic analysis in usability research capturing the latent priorities that structured survey-based quantitative approach may overlook. The emergence of usefulness dimension suggests that clinicians not only sought clarity, trust and a usable system, but also emphasized the practical relevance of model explanations in clinical decision making. Lessons learned from this human-centered AI explainability exercise are generalizable because the clinicians in other health domains also face similar challenges and should be of value to AI researchers and medical informaticians.

[i]END NOTE:
Theoretical foundations of SHAP plot:
The structure of the SHAP plot highlighted how positive and negative contributions add together to reach the final prediction value, f(x).

$$f(x) = E[f(x)] + \sum_{i=1}^{n} \phi_i \quad (1)$$

In Equation (1),
f(x): the model's prediction for a given input x.
E[f(x)]: the expected (or average) value of the model's predictions over the entire dataset. It's often referred to as the "baseline value" or "mean prediction".
$\sum_{i=1}^{n} \phi_i$: the sum of SHAP values, $\phi i$ represents the contribution of feature i to the deviation of f(x) from the baseline value E[f(x)].